\documentstyle[aps,prl,twocolumn]{revtex}

\begin{document}

\draft

\title{Anomalous fluctuations of the condensate in interacting Bose gases} 

\author{S. Giorgini$^{1,4}$, L. P. Pitaevskii$^{2,3,4}$ and S. Stringari$^{4}$}

\address{$^{1}$European Centre for Theoretical Studies
in Nuclear Physics and Related Areas
\protect \\
Villa Tambosi, Strada delle Tabarelle 286, I-38050
Villazzano, Italy} 
\address{$^{2}$Department of Physics, Technion, 32000 Haifa, Israel}
\address{$^{3}$Kapitza Institute for Physical Problems, 117454 Moscow, Russia}
\address{$^{4}$Dipartimento di Fisica, Universit\`a di Trento, \protect\\
and Istituto Nazionale di Fisica della Materia, I-38050 Povo, Italy}

%\date{ }

\maketitle

\begin{abstract}

{\it  We find that the fluctuations of the 
condensate in a weakly interacting Bose gas  
confined in a box of volume $V$ follow the law 
$\langle\delta N_0^2\rangle\sim V^{4/3}$. This anomalous behaviour arises 
from the occurrence of infrared divergencies due to phonon excitations 
and holds also for strongly correlated Bose superfluids.
The analysis is extended to an interacting Bose gas confined in a harmonic 
trap where the fluctuations are found to exhibit a similar anomaly.} 

\end{abstract}


\narrowtext

The recent experimental achievement \cite{EXP} of Bose-Einstein condensation 
(BEC) in dilute atomic vapours
has stimulated a great revival of interest in the theoretical study of this
phenomenon (for un update list of references see \cite{GU}). Among
the several intriguing questions associated with Bose-Einstein condensation
the  problem of fluctuations plays an important role in the more general
context of statistical mechanics (see for example Ref. \cite{ZUK}).

 The fluctuations of the condensate have been studied in a systematic way
in the case of the ideal Bose gas in a box \cite{HAU,FHW,ZUK}, and, 
more recently, 
in
the presence of a harmonic trap \cite{POL,GR,GH}. 
In the ideal Bose gas one finds a
dramatically different behaviour between the canonical and the grand canonical
ensembles.
In particular the results of
Ref. \cite{HAU,ZUK} show that in the canonical ensemble the 
fluctuations of the number of
particles in the ground state of a box of volume $V$ behave as $V^{4/3}$. 
This
law ensures that the relative fluctuations $\langle(\delta N_0)^2\rangle/V^2$ 
of the condensate vanish when $V\rightarrow\infty$,
differently from what happens in the grand canonical ensemble,
where  
$\langle(\delta N_0)^2\rangle = N_0^2 + N_0 \sim V^2$. 
The results obtained in the
canonical ensemble, however, reveal that  
the fluctuations of the condensate are not linear in $V$,
in contrast to the fluctuations of most physical observables. One might
think that this anomalous behaviour is a peculiarity of the ideal Bose gas
and that, as soon as one includes interaction effects, the fluctuations of
$N_0$ take a normal behaviour, linear in $V$. In this letter we address
explicitly this question and we find that, despite the finite compressibility 
of the system, the fluctuations of the condensate still behave as $V^{4/3}$. 

Let us start our investigation in the framework of Bogoliubov theory of
a uniform weakly interacting Bose gas confined in a cubic box of size $L$. 
According to Bogoliubov theory \cite{BOG} the particle 
operator can be expressed in terms of the quasi-particle creation and
annihilation operators as 
$a_{\bf p} = u_p\alpha_{\bf p} + v_p\alpha_{-{\bf p}}^\dagger$,
where ${\bf p}\ne 0$ is the eigenvalue of the momentum operator
${\bf p}=2\pi\hbar{\bf n}/L$, with ${\bf n}=(n_x,n_y,n_z)$. 
The real quantities 
$u_p$ and $v_p$ satisfy the relations
\begin{eqnarray}
&& u_p^2+v_p^2=\frac{\left(\epsilon_p^2+{\rm g}^2n_0^2\right)^{1/2}}
{2\epsilon_p}
\nonumber \\
&& u_pv_p=-\frac{{\rm g}n_0}{2\epsilon_p} \;\;,
\label{uv}
\end{eqnarray}
and, consequently, the normalization condition $u_p^2-v_p^2=1$. 
The energy of the
elementary excitations entering (\ref{uv}) are given by the most 
famous Bogoliubov spectrum
\begin{equation}
\epsilon_p=\left(\left(\frac{p^2}{2m}+{\rm g}n_0\right)^2-
{\rm g}^2n_0^2\right)^{1/2} \;\;,
\label{exen}
\end{equation}
where ${\rm g}=4\pi\hbar^2a/m$ is the coupling constant fixed by the 
s-wave scattering length $a$, 
and $n_0$ is the condensate density which, at low $T$, coincides with the total
density $n=N/V$, apart from higher order effects in ${\rm g}$ arising from   
the quantum depletion of the condensate.
At small $p$ eq. (\ref{exen}) gives the phonon dispersion law 
$ \epsilon_p=cp$ with the sound velocity given by $c=\sqrt{{\rm g}n_0/m}$.
Notice that, due to the discretization of the values of $p$, the phonon 
dispersion law is ensured only if the condition $Na/L\gg 1$ is satisfied.

The total number of particles out of the
condensate is given by the expectation value of the operator
\begin{eqnarray}
N_{out} = \sum_{{\bf p}\ne 0} a_{\bf p}^\dagger a_{\bf p}
&=& \sum_{{\bf p}\ne 0} \Bigl[ u_p^2\alpha_{\bf p}^\dagger\alpha_{\bf p}
+v_p^2\alpha_{\bf p}
\alpha_{\bf p}^\dagger \Bigr.
\nonumber \\
&+& \Bigl. u_pv_p \left(\alpha_{\bf p}\alpha_{-{\bf p}}
+\alpha_{\bf p}^\dagger\alpha_{-{\bf p}}^\dagger\right)
\Bigr]  
\label{Nout}
\end{eqnarray}
and takes the form 
\begin{equation}
\langle N_{out}\rangle=\sum_{{\bf p}\ne 0} \Bigl( (u_p^2+v_p^2)f_p 
+v_p^2\Bigr) \;\;,
\label{Noutav}
\end{equation} 
where
\begin{equation}
f_p=\langle\alpha_{\bf p}^\dagger\alpha_{\bf p}\rangle = 
\left( \exp{\epsilon_p/k_BT}-1 \right)^{-1}
\label{fquasp}
\end{equation}
is the number of elementary excitations present in the system at thermal 
equilibrium. Notice that the number of thermally excited atoms differs from 
the number of elementary excitations because of the factor $(u_p^2+v_p^2)$.  
In the thermodynamic limit ($V\rightarrow\infty$) the sum (\ref{Noutav}) can be 
calculated by integration in momentum space. At low temperature this yields 
the result 
\begin{equation}
\langle N_{out}\rangle = N \left(\frac{m(k_BT)^2}{12nc\hbar^3}
+\frac{8\sqrt{\pi}}{3}\sqrt{a^3n} \right) \;\;,
\label{lowT}
\end{equation}
where $N$ is the total number of particles.
The term proportional to $T^2$ accounts for the phonon
contribution to thermal fluctuations \cite{FER}, while the temperature 
independent term gives the effect of quantum fluctuations \cite{BOG}.  
The depletion of the condensate is obtained by simply taking
the difference $N_0=N-\langle N_{out}\rangle$. 
For the ideal Bose gas confined in a box
[$u_p=1$, $v_p=0$ in (\ref{Noutav})] the same procedure gives the 
well known result $\langle N_{out}\rangle=N(T/T_c)^{3/2}$, where $k_BT_c=
2\pi\hbar^2/m(n/\zeta(3/2))^{2/3}$ is the Bose-Einstein transition 
temperature.  

In a similar way one can proceed to the calculation of the fluctuations of
$N_{out}$. Starting from (\ref{Nout}) one finds
\begin{eqnarray}
\langle\delta N_{out}^2\rangle &=&  
\sum_{{\bf p}\ne 0} \Bigl[ \left( (u_p^2+v_p^2)^2 + 
4u_p^2v_p^2 \right) \left(f_p^2 + f_p\right) \Bigr] 
\nonumber \\
&+& 2\sum_{{\bf p}\ne 0}u_p^2v_p^2 \;\;.
\label{Noutfl}
\end{eqnarray}
To calculate the expectation value of the terms with four quasi-particle 
operators appearing in $\langle N_{out}^2\rangle$ we have used  
Wick's theorem 
\begin{eqnarray}
\langle\alpha_{\bf p}^\dagger\alpha_{\bf p}\alpha_{{\bf p}'}^\dagger
\alpha_{{\bf p}'}\rangle = 
\langle\alpha_{\bf p}^\dagger\alpha_{\bf p}\rangle 
\langle\alpha_{{\bf p}'}^\dagger\alpha_{{\bf p}'}\rangle +
\langle\alpha_{\bf p}^\dagger\alpha_{{\bf p}'}\rangle
\langle\alpha_{\bf p}\alpha_{{\bf p}'}^\dagger\rangle  
\;\;,
\nonumber
\end{eqnarray}
which holds for the creation and annihilation 
operators of   
statistically independent excitations.
In the ideal Bose gas ($u_p=1$, $v_p=0$) result (\ref{Noutfl}) is correct at 
all temperatures 
if one 
works in the grand canonical ensemble, and, as demonstrated in Ref. \cite{ZUK}, 
holds also in the canonical ensemble except near and, of course, above $T_c$.
Physically this behaviour follows from the fact that  
the condensate plays the role of a 
``reservoir'' for the particles out of the 
condensate. On the other hand, in interacting Bose systems  
eq. (\ref{Noutfl}) is valid only at low temperature where 
the number of excitations is small and their interaction can be neglected.
Both in the interacting and non-interacting case eq. (\ref{Noutfl})  
allows us to calculate 
the fluctuations of the condensate in the canonical ensemble where one has 
trivially $\langle\delta N_0^2\rangle =\langle\delta N_{out}^2\rangle$ 
\cite{N2}.   

In equation (\ref{Noutfl})  
one immediately points out the occurrence of a strong infrared
divergency arising from the term proportional to $f_p^2$. 
In fact at low $p$ one has 
$u_p^2\simeq v_p^2\simeq {\rm g}n_0/2cp$ and $f_p\simeq k_BT/cp$.  
This results in a $1/p^4$ infrared divergency yielding 
a divergent integral in 3D. Also in the ideal gas one finds
a similar behaviour. In fact in this case $u_p^2=1$, $v_p^2=0$, while 
$f_p\simeq 2mk_BT/p^2$.
The occurrence of this divergency implies that the sum (\ref{Noutfl}) 
cannot be
calculated by integration in momentum space, but is dominated by the  
discretized sum over the low-energy phonon modes.  
The final result for the leading terms in the large-$N$
limit is given by
\begin{eqnarray}
\langle\delta N_0^2\rangle &=&  A \left(\frac{mk_BT}{\hbar^2}\right)^2
V^{4/3} 
\nonumber \\
&+& \frac{1}{3\pi^2}\frac{k_BTm^2c}{\hbar^3} V\log\left(\frac{Vm^3c^3}
{\hbar^3}\right) + o(V) \;\;,
\label{diver}
\end{eqnarray}
where $A=2/(\pi)^4 \sum_{{\bf n}\ne 0} 1/{\bf n}^4=0.105$.
The term containing $\log{V}$ arises from the contribution linear 
in $f_p$ of eq. (\ref{Noutfl}) which yields an integral in momentum space 
exhibiting a logarithmic divergency.
It is worth noting that the coefficient of the
leading term does not depend on the velocity of sound (and hence on the
interaction coupling constant), nor on the number of particles. In the
ideal  Bose gas (IBG) using cyclic boundary conditions the result looks similar, 
but the coefficient differs by a
factor 2 \cite{N1} 
\begin{equation}
\langle\delta N_0^2\rangle_{C}^{IBG} = 2 A \left(\frac{mk_BT}{\hbar^2}\right)^2
V^{4/3} + o(V) \;\;.
\label{IBG}
\end{equation}
In this case it is important to specify that 
the result holds in the canonical (C)
ensemble.
The IBG result (\ref{IBG}) holds if the condition $Na/L\ll1$ is satisfied.

The anomalous behaviour (\ref{diver}) exhibited by the fluctuations of the 
condensate reflects the occurrence of infrared divergencies in the  
one-particle Green's function \cite{GF}.  
These phenomena are well 
accounted for by the hydrodynamic theory of Bose superfluids (see for example 
Ref. \cite{GPS}). 
Actually this theory can be directly employed to derive result (\ref{diver})
which consequently holds not only in the case of a weakly-interacting Bose
gas, but also in strongly-interacting Bose fluids, like liquid $^4$He.
Notice that in this case the condensate fraction at $T=0$ sizably differs 
from the ideal gas value and that the sound velocity $c$, entering the $\log V$ 
term in eq. (\ref{diver}), cannot be calculated using Bogoliubov theory.  

The fluctuations of the condensate can also be evaluated at $T=0$ 
[$f_p=0$ in eq. (\ref{Noutfl})]. In this case the sum has no infrared  
divergencies and can be safely calculated through direct integration in 
momentum space. This yields the result
\begin{equation}
\langle\delta N_0^2\rangle (T=0) =  
\sqrt{\pi} (an_0)^{3/2} V \;\;,
\label{T0}
\end{equation}
revealing that, at $T=0$, the fluctuations of the condensate have a 
``normal''  
behaviour, i.e. they are linear in $V$. 
These fluctuations are however large compared to the ones
of the total number of particles, which in the grand canonical 
ensemble behave as $\langle\delta N^2
\rangle\simeq V^{2/3}\log V$. This latter result is obtained by writing 
\begin{equation}
\langle\delta N^2\rangle = n \int d{\bf r}_1 d{\bf r}_2 \nu(|{\bf r}_2 -
{\bf r}_1|) \;\;,
\label{delN}
\end{equation}
where $\nu(r)$ is the density-density static correlation function
$n \nu(|{\bf r}_2-{\bf r}_1|) \equiv \langle\delta n({\bf r}_2)\delta 
n({\bf r}_1)\rangle$.
In an interacting Bose superfluid at $T=0$ the large $r$ behaviour of $\nu(r)$ 
is fixed by the asymptotic law \cite{LP}
\begin{equation}
\lim_{r\rightarrow\infty} \nu(r) = - \frac{\hbar}{2\pi^2mc}\frac{1}{r^4} \;\;.
\label{larger}
\end{equation}
Let us calculate the fluctuations of $N$ in a sphere of radius $R\gg \hbar/mc$.
The domain of integration in (\ref{delN}) is then $r_1$,$r_2\le R$ and 
one can write 
\begin{equation}
\langle\delta N^2\rangle = 4\pi n\int_0^{2R} dr \; r^2\tau(r)\nu(r) \;\;,
\label{delN1}
\end{equation} 
where $\tau(r)=2\pi R^3(2-3r/2R+r^3/8R^3)/3$ is the overlapping volume of two 
spheres of radius $R$ whose centers are separeted by the distance $r$.
The integral of the $r$ independent term in $\tau(r)$ gives a contribution 
proportional to $R^2$ as can be seen by exploiting the sum rule 
$4\pi\int_0^\infty dr \;r^2\nu(r)=S(k=0)=0$,
holding for systems where the static form factor $S(k)$ vanishes in the 
long-wavelength limit. The dominant contribution to $\langle\delta N^2\rangle$
comes thus from the term linear in $r$ in $\tau(r)$. One finally finds 
the relevant
result \cite{N3} 
\begin{equation}
\langle\delta N^2\rangle (T=0) = 
\frac{2n\hbar}{mc} R^2\log\left(\frac{Rmc}{\hbar}
\right) \;\;,
\label{delN2}
\end{equation}
holding with logarithmic accuracy.
    
The analysis developed above can be easily extended to the case of a Bose gas 
confined in a harmonic potential of the form 
$V_{ext}=m\omega_0^2r^2/2$,  
where, for simplicity, we have considered isotropic 
trapping. 
Also in this case one can define a 
proper 
``thermodynamic limit''.  
This is achieved by letting 
$\omega_0\rightarrow 0$, $N\rightarrow\infty$ while keeping $\omega_0^3N$ fixed
\cite{US,DAM}.   
Differently from the gas confined in a box, in the presence of harmonic 
trapping the density of the system is inhomogeneous and it is convenient
to use
the following decomposition for the particle field operator
\begin{equation}
\psi({\bf r}) = 
\Phi({\bf r}) + \sum_i \Bigl( u_i({\bf r})\alpha_i + v_i^\ast({\bf r})
\alpha_i^\dagger \Bigr) \;\;,
\label{decomp1}
\end{equation}
where $\Phi({\bf r})=\langle
\psi({\bf r})\rangle$ is the order parameter and the index $i$ labels the 
elementary excitations of the system for which the normalization condition 
$\int d{\bf r}(|u_i|^2-|v_i|^2)=1$ holds.  
The number of particles out of the condensate is given by the operator 
$N_{out}=\int
d{\bf r} (\psi^\dagger({\bf r})-\Phi^\ast({\bf r}))(\psi({\bf r})-
\Phi({\bf r}))$.

Let us first discuss the non-interacting gas. In this case the fluctuations 
of $N_{out}$ can be written as  
$\langle\delta 
N_{out}^2\rangle = \sum_if_i(f_i+1)$, where $i$ labels the single-particle 
excited states of the oscillator Hamiltonian.  
By using the semiclassical approximation for these
states \cite{US}, we obtain the result \cite{POL} 
\begin{equation}
\langle\delta N_0^2\rangle_{C}^{IBG} = 
\frac{\pi^2}{6\zeta(3)}N\left(\frac{T}{T_c}\right)^3 
\label{trap}
\end{equation}
for the fluctuations of the condensate 
in the canonical ensemble.  
Result (\ref{trap}) holds for temperatures smaller than the critical 
temperature for BEC \cite{IGAS} 
\begin{equation}
k_BT_c=\hbar\omega_0\left(\frac{N}{\zeta(3)}\right)^{1/3}
\label{Tc}
\end{equation}
which is kept fixed in the thermodynamic limit. 
The applicability of the
semiclassical approximation is ensured for temperatures much larger than
the oscillator temperature ($k_BT\gg\hbar\omega_0$) and is a consequence of the
fact that the sum $\sum_i f_i(f_i+1)$ takes most of its contribution from 
excitations
with energy $\sim k_BT$.
The changes in the fluctuations of the condensate (\ref{trap}) due to the 
use of the microcanonical ensemble can be also calculated and result in a 
change of the numerical prefactor \cite{GR,GH}.
 
Result (\ref{trap}) shows that the fluctuations of the condensate in a harmonic
trap have a normal behaviour, i.e. they are proportional to $N$. 
This behaviour differs from result 
(\ref{IBG}) for the 
ideal Bose gas in a box where the fluctuations have a stronger dependence 
on $N$ (or $V$).
This fact is not a surprise because it is well known that confinement
reduces the effects of thermal fluctuations.  For example,  the
2D ideal Bose gas in a harmonic trap exhibits Bose-Einstein condensation at
finite temperature, differently from what happens in the uniform gas where
thermal fluctuations have a destabilizing effect on the condensate.

Interactions introduce however a deep difference in
the above behaviour. In fact for repulsive forces  
the system exhibits a phonon-type
behaviour at low excitation energies. This suggests that the fluctuations of the condensate
will be dominated by the discretized low energy modes which in the 
large-$N$ limit 
obey the dispersion 
law \cite{STR} 
\begin{equation}
\epsilon(n,\ell)=\hbar\omega_0 \left( 2n^2+2n\ell+3n+\ell\right)^{1/2} \;\;.
\label{Strexc}
\end{equation}
These phonon-like collective excitations have been recently  
measured \cite{EXP1} and the observed frequencies are in excellent
agreement with theory. 
Let us recall that the  
dispersion law (\ref{Strexc}) is valid if the conditions $Na/a_{HO}\gg1$ and 
$\epsilon(n,\ell)\ll\mu$ are satisfied. 
Here   
$a_{HO}=\sqrt{\hbar/m\omega_0}$ is the oscillator length and $\mu$ is the 
chemical potential calculated at zero temperature. 
The condition $Na/a_{HO}\gg1$, which is well satisfied in the recent 
experiments on
BEC,  
is the analogue 
of the condition $Na/L\gg 1$ which must be  
satisfied in the transition from the ideal to the Bogoliubov gas in the box. 
Furthermore, the above condition yields the 
``Thomas-Fermi''  
result for the ground state density of the system $n_0({\bf r})=
|\Phi({\bf r})|^2=\mu/{\rm g}(1-r^2/R_c^2)\theta(R_c-r)$, where 
$R_c=2\mu/(m\omega_0^2)$ is the radius of the condensate and $\mu=
\hbar\omega_0(15Na/a_{HO})^{2/5}/2$.
In the same limit one can
also calculate the quantities $u_i$ and $v_i$ of (\ref{decomp1}) relative 
to the low-lying excitations (\ref{Strexc}). One obtains the   
following leading behaviour 
\cite{GRI} 
\begin{equation}
u_i({\bf r})\simeq -v_i({\bf r})\simeq\sqrt{\frac{{\rm g}n_0({\bf r})}
{2\epsilon_i}}
\chi_i({\bf r}) \;\;,
\label{uvhydr}
\end{equation}
where $\chi_i({\bf r})$ 
is the velocity potential associated with the 
collective mode satisfying the condition $\int d{\bf r} 
\chi_i^\ast({\bf r})
\chi_j({\bf r})
=\delta_{ij}$. After a straightforward calculation 
one finds that the leading term in the 
fluctuations of the condensate exhibits an anomalous behaviour proportional to 
$N^{4/3}$:
\begin{equation}
\langle\delta N_0^2\rangle =  
B \;\frac{T^2}{T_c^2} \left(
\frac{ma^2k_BT_c}{\hbar^2}\right)^{2/5} N^{4/3} , 
\label{trap1}
\end{equation}
where $T_c$ is given in (\ref{Tc}) and $B$ is a dimensionless coefficient  
\begin{eqnarray}
B &=& \gamma \sum_{nn'\ell}\frac{(2\ell+1)}
{(2n^2+2n\ell+3n+\ell)(2n'^2+2n'\ell+3n'+\ell)}
\nonumber \\
&\times& 
\frac{\left( \int_0^1 dx x^{2(\ell+1)} (1-x^2) 
P_\ell^{(2n)} P_\ell^{(2n')}
\right)^2} {\int_0^1 dx x^{2(\ell+1)}(P_\ell^{(2n)})^2
\int_0^1 dx x^{2(\ell+1)} (P_\ell^{(2n')})^2},
\label{Bcoef}
\end{eqnarray}
with $\gamma=15^{4/5}/(2(\zeta(3))^{8/15})$.  
In the above equation $n,n',\ell=0,1,2,..$ and $\ell\neq 0$ if $n\,(n')\,=0$, 
while $P_\ell^{(2n)}\equiv 
P_\ell^{(2n)}(x)$ are 
the polinomials introduced in Ref. \cite{STR}. 
It is worth noting that the inclusion of interaction  
reproduces a
behaviour similar to the one exhibited by the gas confined in a box, i.e.  
proportional to
$N^{4/3}$. This result reveals that the apparently similar behaviour in the
ideal  and interacting Bose gases in a box is accidental and that 
the physical effects induced by interactions show up in a clearer way in the
presence of the harmonic trap. 

The consequences of interactions on the fluctuations of the condensate are
expected to be even more dramatic in the case of attractive forces, where
for values of $N$ close to the critical size for the onset of instability,
the monopole vibration has  a
vanishing frequency with consequent major contribution to the sum 
(\ref{trap1}).
This situation, which is characterized by a full instability of the 
grand canonical 
ensemble, will be considered in a subsequent work.

It is finally important to keep in mind that only the fluctuations of the 
condensate 
exhibit the anomalous behaviour discussed in the present work. 
In particular the 
fluctuations of the density operator $\psi^\dagger({\bf r})\psi({\bf r})$ 
are expected to exhibit a normal 
behaviour in both
uniform and non uniform interacting Bose gases.


\begin{references}
\bibitem{N2} The distinction between canonical and  
grand canonical ensemble for the calculation of $\langle\delta N_0^2\rangle$ 
is crucial only in the ideal Bose gas. The presence of interactions 
stabilizes the fluctuations of $N$ and ensures the equivalence, in the 
thermodynamic limit, of the two ensembles also for the low-$T$ behaviour  
of $\langle\delta N_0^2\rangle$.
\bibitem{N1} In the case of the ideal Bose gas one can easily work 
also with different boundary conditions \cite{ZUK}.
\bibitem{N3} For an ideal gas in the grand canonical ensemble the same method 
gives $\langle\delta N^2\rangle (T=0)=\langle N\rangle$, as can be directly
obtained from eq. (\ref{delN}) with $\nu({\bf r})=\delta({\bf r})$.
\end{references}
\end{document}